# Self-assembled fluorescent nanodiamond layers for quantum imaging

Katherine Chea[1], Erin S. Grant [1], Kevin J. Rietwyk[1], Hiroshi Abe[2], Takeshi Ohshima[2,3], David A. Broadway[1], Jean-Philippe Tetienne[1], Gary Bryant[1], Philipp Reineck[1]

[1] School of Science, RMIT University, Melbourne, VIC 3001, Australia
[2] National Institutes for Quantum and Radiological Science and Technology, Takasaki, Gunma, 370-1292, Japan
[3] Department of Materials Science, Tohoku University, Aoba, Sendai, Miyagi 980-8579, Japan

**Abstract.** The nitrogen-vacancy (NV) center in diamond is emerging as a powerful tool for imaging magnetic and electric signals at the microscale and below. However, most imaging demonstrations thus far have relied on costly, millimeter-sized bulk diamond substrates, which cannot be easily scaled or integrated with other materials. Here, we report a scalable method for fabricating NV-containing dense and homogenous fluorescent nanodiamond (FND) layers through electrostatic self-assembly and demonstrate the utility of the FND layers for magnetic imaging. We investigate the effect of FND concentration in suspension, substrate immersion time, and solvent pH on the FND density on the substrate. We identify optimized self-assembly conditions that maximize the FND density while minimizing aggregation. Using FND layers on a quartz substrate, we demonstrate magnetic field and magnetic noise imaging at the microscale, based on NV optically detected magnetic resonance magnetometry and $T_1$ relaxometry, respectively. Our results provide a direction for the development of cost-effective and scalable FND layers and surface coatings. This paves the way for on-demand quantum sensing and imaging on a broad range of surfaces based on NV centers and other diamond quantum emitters.

**Introduction.** Fluorescent color centers are proving to be powerful nanoscale probes for sensing and imaging magnetic and electric fields, as well as temperature, with sub-micron spatial resolution. The most commonly used and technologically advanced among them is the nitrogen vacancy (NV) center in diamond. In the context of quantum imaging, the NV center has been used to image 2D materials [1–6] and biominerals [7] based on the NV spin properties. Most imaging approaches reported to date utilize expensive, millimeter-scale bulk diamond substrates, which are difficult to manufacture at the centimeter scale and above and are challenging to integrate into other materials.

By contrast, FNDs containing NV centers are much less expensive, are already produced at scale commercially, and can be integrated into other materials. However, for imaging and sensing applications, positioning and arranging FNDs at the nanoscale near the sensing target is non-trivial. Many approaches have been explored for different applications, including molecular targeting in cells and assays [8–10], optical tweezers [11–13], FND-doped bioscaffolds [14] or flexible membranes[15], FND DNA origami [16], and creating FND aggregates on surfaces [17–21].

However, to use FNDs for quantum imaging with microscale or higher spatial resolution, dense FND layers on a solid substrate are needed. These FND layers could be created on a broad range of rigid and flexible substrates and act as a functional coating that enables NV-based quantum sensing and imaging. These could be used for magnetic imaging at the microscale in cases where bulk diamond chips are too expensive or challenging to interface with the sample, for sensing chips for diagnostic devices where dense FND monolayers potentially offer higher sensitivities, or even for biomedical implants with quantum sensing functionality. Yet to date, few studies have explored the creation of dense, uniform FND layers, which will be discussed in more detail below.

Nanodiamond (ND) layers of different densities have been extensively explored in the context of the growth of single-crystal diamond using chemical vapor deposition[22] with many different approaches, including electrostatic self-assembly[23]. These generally use detonation nanodiamonds (DNDs) for their small primary particle size of ~ 5 nm and near spherical particle shape. However, the controlled engineering of color centers like the NV center in DNDs remains challenging, and DNDs, as a result, are not commonly used for quantum sensing applications. Instead, commercially available high-pressure high-temperature NDs are generally employed that either already contain high densities of NV centers or can be processed to host NV centers at concentrations of 1-10 ppm[24] These are often referred to as FNDs and their physical, chemical, colloidal, and optical properties greatly differ from those of DNDs. Hence, protocols developed for DNDs cannot be directly applied to FNDs.

Past studies on the creation of FND layers have mainly focused on the creation of controlled positioning of individual particles into different types of sparse FND arrays[15,25–29]. For example, Kianinia et al have used covalent binding of carboxylated FNDs to a patterned silicon wafer substrate via a carbodiimide crosslinker to create FND arrays with a >90% yield[26]. Members of our team have created similar arrays using a patterned photoresist[25]. More recently, Shulevitz et a have demonstrated another template-based approach to create similar arrays[28]. Andrich et al have created a polymer-based FND sensing array with tunable array spacings on the micron scale for temperature sensing[15]. Foy et al[19] have demonstrated magnetic field and temperature imaging using drop-cast and randomly deposited FNDs with a highly inhomogeneous particle distribution. However, for quantum imaging and sensing applications, dense and homogenous FND layers offer the highest spatial imaging resolution and NV signal per unit surface area.

Here, we report a scalable and straightforward method for creating dense FND layers on solid substrates, based on the electrostatic self-assembly of negatively charged FNDs onto positively charged substrates as illustrated in Figure 1 a) and b). A glass or silicon substrate is functionalized with a positively charged polymer (polyallylamine hydrochloride, PAH) and immersed in an aqueous FND suspension, allowing FNDs to attach to the surface electrostatically. The resulting FND layers exhibit homogenous NV PL on a millimeter scale and can be used for a range of imaging and sensing applications, including magnetic, temperature, and NV charge state-based imaging and sensing. We investigate the effects of suspension FND concentration, pH and ionic strength as well as substrate immersion time, on the resultant FND substrate density. We identify optimized self-assembly parameters that maximize the density of single FNDs on the surface and avoid aggregate formation. We demonstrate the utility of the FND layers for magnetic imaging by depositing magnetic $Fe_2O_3$ particles on the FND-coated substrates (Figure 3c) and image the $Fe_2O_3$ particles' magnetic properties using NV optically detected magnetic resonance (ODMR) spectroscopy and NV $T_1$ spin relaxometry. Our results aid the development of inexpensive and scalable quantum sensing and imaging substrates based on dense FND layers containing NV centers and other emerging color centers.

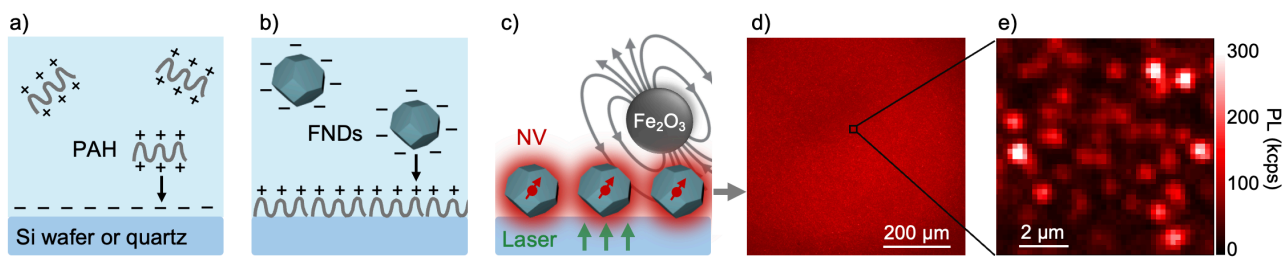

**Figure 1**. Schematic illustration of the fabrication process and PL images of the self-assembled fluorescent nanodiamond layers. a) A Si, quartz, or glass substrate is immersed in an aqueous solution of polyallylamine hydrochloride (PAH), which attaches to the substrate to create a positively charged surface. b) The PAH functionalized substrate is immersed in an FND suspension and negatively charged FNDs electrostatically attach to the glass substrate. c) PL from NV centers in the FNDs can be used to image the magnetic properties of materials on the surface. An iron oxide particle is shown as an example. d) Wide-field PL image of an FND layer showing homogeneous red NV PL over a 630 × 630 μm area of the substrate. e) High-resolution confocal PL image of a small region on the substrate showing individual FNDs.

**Results**. To maximize the density of individual FNDs on the substrate surface, two processes must be balanced while avoiding FND aggregation: (1) electrostatic attraction between the positively charged substrate surface and negatively charged FNDs; and (2) electrostatic repulsion between FNDs. (1) is the driving force for FND surface attachment and (2) stabilizes FNDs in suspension and prevents aggregation. At the same time, FND-FND repulsion (2) also limits the FND density on the surface: an FND attached to the surface will prevent other FNDs from attaching to the substrate in its immediate vicinity via electrostatic repulsion. Hence, to maximize the FND density on the surface, FND-FND repulsion must be reduced, which increases the probability of aggregate formation. The simplest way to reduce electrostatic repulsion is electrostatic shielding via the addition of salt. However, this will also reduce the attraction between FNDs and the surface. Increasing the FND concentration in suspension and the substrate immersion time do not directly affect FND-FND repulsion. However, both increase the probability for FND surface attachment due to higher FND diffusion towards the surface per unit time, and by allowing more time for FNDs to diffuse towards the surface, respectively. However, both may also lead to more aggregation. One parameter that may affect oppositely charged surfaces differently is pH. In our case, the amine that confers the positive charge to PAH has a pKa above 8. Hence, reducing the pH from 7 to 4, for example, will not affect the protonation and overall charge of PAH. The pKa of oxygen containing groups on the FND surface like carboxylates are in the range of 3-5 depending on the exact molecular or atomic environment [30]. Hence, a reduction in pH may lead to an increase in the protonation of FND surface groups and a decrease in the FND surface charge. Therefore, we explored the effect of FND concentration in suspension, substrate immersion time in the FND suspension, and solution ionic strength and pH on FND density on the substrate surface.

Figure 1a and 1b show the two main steps involved in the FND self-assembly. Silicon wafers were used in the first part of the study due to their low surface roughness, which enabled a reliable AFM-based analysis of the FND surface coverage. Quartz substrates were used for quantum imaging experiments since these allow for imaging through the substrate and exhibit low background PL. To fabricate FND layers, the substrate was cleaned, immersed in an aqueous solution of PAH (1 mg mL$^{-1}$), and then immersed in aqueous FND suspensions of varying FND concentrations, pH levels, and ionic strengths for durations ranging from 1 to 1000 s. See methods section and SI for more details on the sample preparation.

Figures 1d) and 1e) show PL images of a 630 μm × 630 μm and ~9 μm × 9 μm area of a resulting sample with a high density of FNDs, acquired using a commercial wide field fluorescence microscope. The lower magnification image (Figure 2d) shows homogeneous FND PL from NV centers across the entire field of view. The zoomed image in Figure 2e) shows that the PL originates from individual FNDs or small FND aggregates.

We used AFM imaging to study the effect of FND concentration, substrate immersion time, and pH and ionic strength of the FND suspension on the FND density on the substrate surface. In a typical experiment, a 50 μm × 50 μm image was acquired (see SI Figure S1 for full images) and analyzed using semi-automated image analysis using Gwyddion and IgorPro to determine the number of FNDs attached to the surface per 5 μm × 5 μm. Using a combination of dynamic light scattering data and AFM images of isolated FNDs, we determined a z-height threshold of 55 nm to separate single FNDs from FND aggregates and stacked FNDs on the substrate. The FNDs have a hydrodynamic diameter of 120 nm. Due to their disk-like shape, particles will typically attach to the substrate with one of their larger faces and show a z-height in AFM images that is well below their DLS hydrodynamic diameter [31]. See SI Figure S2 for a typical AFM image and a DLS particle size distribution.

Figures 2a) to 2d) show typical AFM z-height images of FNDs on a Si wafer for suspension FND concentrations from 0.01 to 10 mg mL$^{-1}$, for a fixed substrate immersion time of 100 seconds. The images qualitatively show an increasing FND density from 0.01 to 1 mg mL$^{-1}$. For 10 mg mL$^{-1}$, the density decreases, and the relative occurrence of aggregates increases. Figure 2f) is a quantitative analysis of these trends and shows the average number of FNDs per 5 μm × 5 μm area for single FNDs and FND aggregates as a function of FND concentration in suspension during the self-assembly for an immersion time of 100 s. The error bars represent the standard deviation across four areas (25 μm × 25 μm) analyzed individually across a total area of 50 μm × 50 μm. Both single FND and the FND aggregate density increase from 0.01 to 1 mg mL$^{-1}$ and decrease again at 10 mg mL$^{-1}$, with a clear peak at 1 mg mL$^{-1}$.

Figure 2g) shows the average FND surface-to-surface distance ($d_{avg}$) as a function of FND concentration for all FNDs on the substrate surface, assuming disk-shaped particles with a diameter of 120 nm in the x-y imaging plane (see SI for details on the calculation). The horizontal dashed blue line indicates the optical diffraction limit of 290 nm for green light (520 nm) and an objective with NA 0.9. At a concentration 1 mg mL$^{-1}$, $d_{avg}$ reaches the diffraction limit, suggesting that at this concentration the NV PL can be collected from the entire substrate surface.

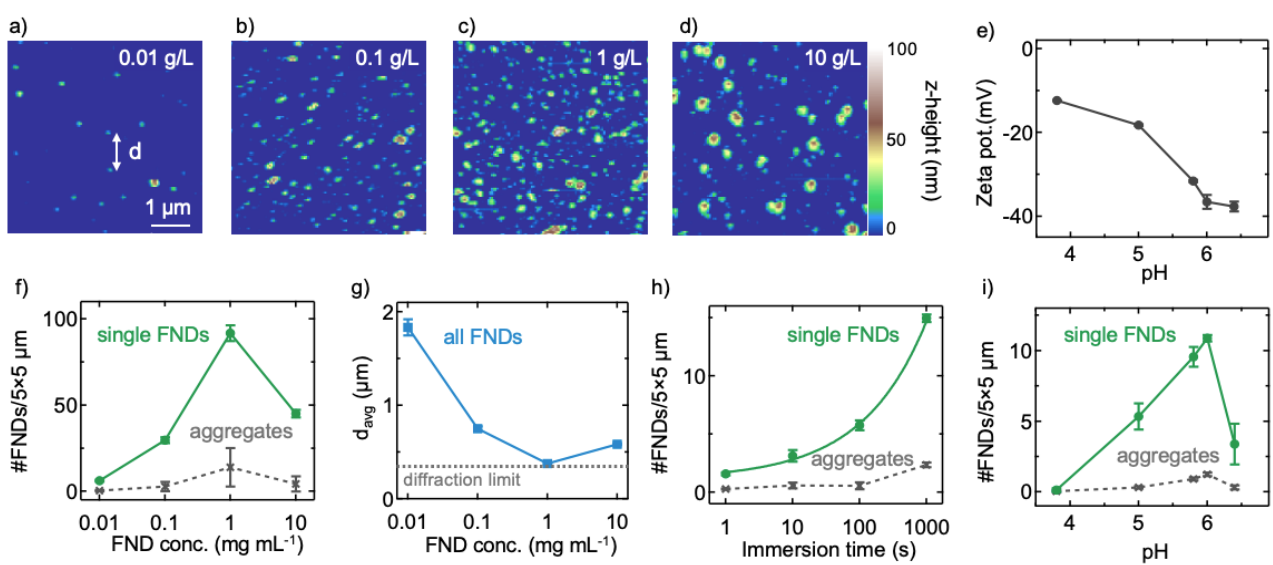

**Figure 2**. The effect of FND concentration in suspension, substrate immersion time and suspension pH on the FND surface coverage. a) to d): AFM images (5×5 μm) of FND layers assembled from suspensions containing different concentrations of FNDs as indicated in each image. e) FND zeta potential as a function of pH. f) Average number of single (green trace) and aggregated (grey dashed trace) FNDs attached to a 5×5 μm area on the substrate surface as a function of the FND concentration in suspension (100 s immersion time). g) Average surface-to-surface distance of FNDs (single and aggregated FNDs) attached to the substrate surface as a function of the FND concentration in suspension. The horizontal dashed blue line indicates the optical diffraction limit. h) Average number of FNDs attached to the substrate surface as a function of substrate immersion time in the FND suspension (FND concentration 0.1 mg mL$^{-1}$). i) Average number of FNDs attached to the substrate surface as a function of pH in the FND suspension.

Figure 2h) shows the number of FNDs attached to the surface as a function of substrate immersion time for an FND concentration of 0.1 mg mL$^{-1}$ in suspension. The number of single FNDs attached to the surface increases with the square root of the substrate immersion time (solid green trace in Figure 2h), while the number of FND aggregates only shows a significant increase after 1000 s (~16 min).

Figures 2e) and 2i) show the zeta potential of the FNDs in suspension and the number of FNDs attached to the substrate as a function of suspension pH, respectively, for an FND concentration of

0.1 mg mL$^{-1}$ in suspension. The pH was controlled by the addition of HCl (see SI Table S1 for details), and both figures show measured pH values. At pH 6.4 (without HCl added), the particles exhibit a strong negative zeta potential of -38 mV, which increases continuously with decreasing pH to -12 mV at pH 4. In the same pH range, the number of FND attached to the surface shows a pronounced increase from pH 6.4 to pH 6.0 and 5.8 and then decreases to zero attached particles at pH 4. We also investigated the effect of the ionic strength of the solvent on FND zeta potential and the number of FNDs attached to the surface via the addition of NaCl (see SI Figure S3). While the zeta potential increased from -43 mV at 10 µM NaCl to -31 mV at 10 mM NaCl due to electrostatic shielding, we did not find a systematic change in the FND density on the substrate surface with the change in ionic strength.

Qualitatively, the results shown in Figure 2 can be understood in terms of the two processes introduced earlier: (1) attraction between the substrate surface and FNDs; and (2) repulsion between FNDs that stabilizes the particles in suspension. Increasing FND concentration and substrate immersion time both increase the total number of diffusion-driven direct FND encounters with the substrate surface. In the simplest model, FND attachment is irreversible, and once an FND has attached to the surface, this attachment site is no longer available. Assuming spherical particles, this process is described by the simplest random sequential adsorption model[32]. While this model is too simple to describe the system studied here, it qualitatively explains the saturation behavior seen in Figures 2f) and 2h) (see SI Figure 4 for linear x-axis scaling).

However, we find that increasing the concentration above 1 mg mL$^{-1}$ does not increase the number of FNDs attached to the surface, potentially due to increased FND aggregation in suspension. For any FND concentration, on the timescale of hours, sedimentation begins to affect FNDs with diameters greater than 100 nm. The immersion time that maximizes the density of single FNDs for the FND concentration of 0.01 mg ml$^{-1}$ is likely above 1000 s, but at the expense of additional aggregates. The addition of 1-10 µM of HCl increases the number of FNDs attached to the surface, likely due to a small decrease in FND-FND repulsion indicated by a reduction in FND surface charge. At lower pH levels, FNDs become colloidally less stable, resulting in a reduction in FND density on the substrate.

Based on the results in Figure 2, we then fabricated optimized dense FND layers on quartz substrates to investigate the utility of the FND layers for magnetic imaging. We investigated 1 mg mL$^{-1}$ FNDs for an immersion time of 1000 s with and without the addition of 1 µM HCl. We found that the addition of HCl slightly decreases the FND density on the surface for this FND concentration in suspension and immersion time, and hence identified 1 mg mL$^{-1}$ FNDs and 1000 s substrate immersion time without HCl as the optimum self-assembly conditions. AFM and confocal PL images of the resulting FND layers are shown in SI Figure S5. The FND surface coverage determined from the AFM image was 16.8%, which is a significant improvement compared to the highest FND coverage of 11.5% obtained for this concentration with 100 s immersion time (Figures 2c and 2f). While this surface coverage is well below a dense monolayer coverage, a diffraction-limited confocal PL image shows that it is sufficient to collect strong NV PL (> 20 kcps on an avalanche photodiode) from 91.4% of the substrate surface. More than 10k photocounts of NV PL can be detected from 99.9% of the substrate surface, which is about an order of magnitude above the dark counts of 1 kcps of our detector. See SI Figure S5 for an AFM image and confocal PL image of the substrate.

To investigate magnetic field and magnetic noise imaging with FND layers, we used a custom-built wide-field quantum diamond microscope [33] to image FND layers before and after the deposition of magnetic $Fe_2O_3$ micro- and nanoparticles. Static magnetic fields induce Zeeman splitting of the NV spin sublevels, resulting in a broadening of the ODMR spectrum of FNDs, whose crystal axes are randomly oriented. This broadening is proportional to the strength of the local magnetic field as discussed in more detail below. In contrast, fluctuating magnetic fields—particularly those resonant with the NV spin transitions—induce spin relaxation, shortening the NV $T_1$ time, making it a sensitive probe of fluctuating nanoscale magnetic fields.

Figure 3a) shows a bright-field reflectivity image of the FND layer on a quartz substrate before (inset) and after drop-casting of $Fe_2O_3$ particles. While individual FNDs cannot be seen due to their small size and homogeneous coverage across the surface, smaller and larger $Fe_2O_3$ particle aggregates can be observed across the field of view. We then collected NV optically detected magnetic resonance (ODMR) spectra and $T_1$ spin relaxation decay traces for each pixel in the image, both before and after the deposition of $Fe_2O_3$ particles, to investigate their magnetic properties.

Figures 3 b) and 3c) show typical ODMR spectra and $T_1$ decay traces, respectively. To acquire the ODMR spectra and $T_1$ decays, the FND PL signal from ~8 × 8 μm regions of interest (ROI, see SI Figure S6 for details) on the substrate was analyzed before and after $Fe_2O_3$ particle deposition. In the absence of $Fe_2O_3$ particles, the FND NV ODMR spectrum shows a characteristic resonance at a MW frequency just below 2.87 GHz with a full-width half maximum (FWHM) of 23 MHz, consisting of two smaller dips split by 9 MHz due to strain and charged defects in the FNDs. In the presence of the magnetic $Fe_2O_3$ particles, the FWHM increases to 36 MHz and the center frequency shifts to 2.86 GHz, likely due to light-induced heating of the $Fe_2O_3$ particles. See SI Figure S7 a) for a map of the resonance frequency converted to a light-induced temperature change.

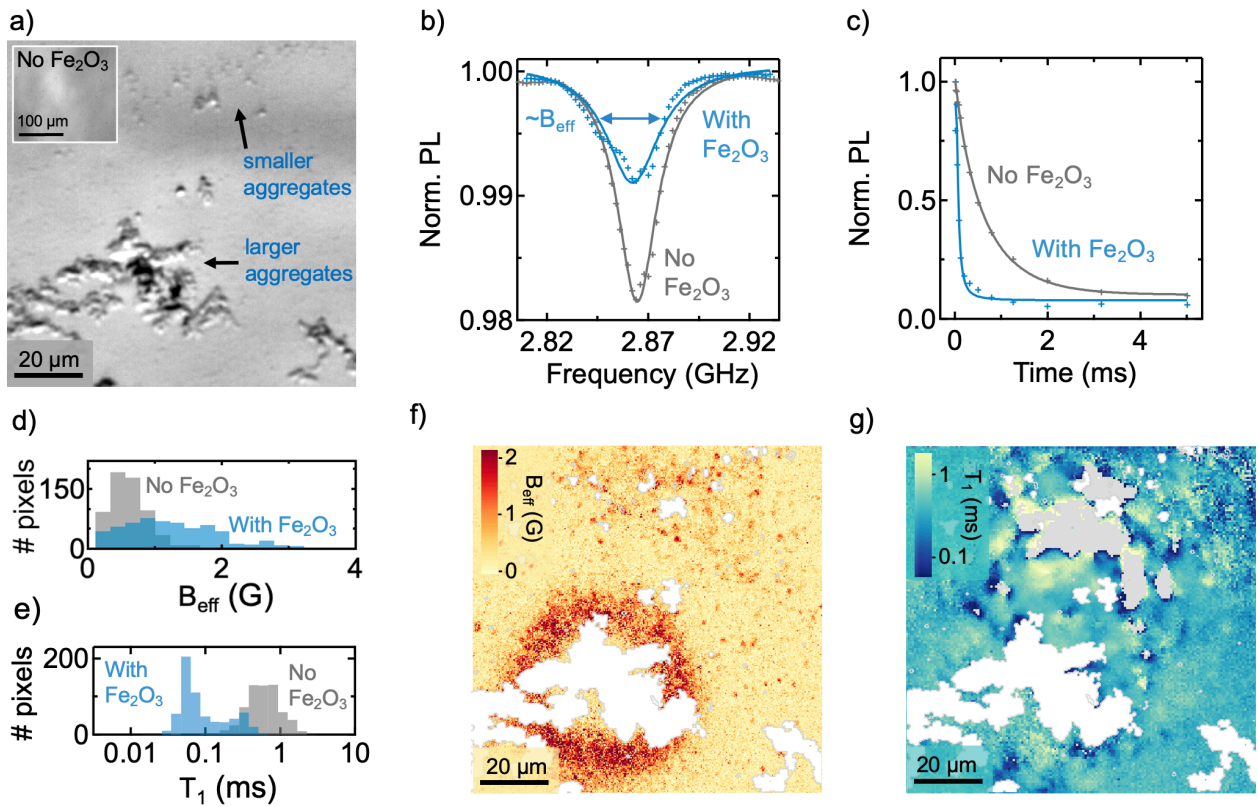

**Figure 3**. Demonstration of magnetic field and magnetic noise imaging using self-assembled FNDs. a) Bright-field image of $Fe_2O_3$ NPs deposited on a FND NP layer on a quartz substrate. The inset shows the substrate before deposition of $Fe_2O_3$ NPs. b) Typical NV ODMR spectra before (grey trace) and after (blue trace) deposition of $Fe_2O_3$ NPs. c) Typical NV $T_1$ decay traces before (grey trace) and after (blue trace) deposition of $Fe_2O_3$ NPs. d) & e) Histogram of effective magnetic field values, $B_{eff}$, and $T_1$ relaxation times determined from individual ODMR spectra (d) and $T_1$ decay traces (e) acquired for 625 pixels over an 8.1×8.1 μm area on the substrate. f) & g) Maps of $B_{eff}$ (f) and $T_1$ times (g) after the deposition of $Fe_2O_3$ NPs. Regions covered by $Fe_2O_3$ NPs are shown in white and areas where $T_1$ decay times could not be determined in grey.

We can estimate a relative local magnetic field ($B_{eff}$) from the broadening of the ODMR FWHM, which we define according to the following equation:

$$B_{eff} = \frac{FWHM}{2\gamma_{NV}} \quad (1)$$

where the FWHM is determined based on a single Lorentzian fit to the ODMR spectrum and $\gamma_{NV}$, is the gyromagnetic ratio of the NV center. Figure 3d) shows a histogram of $B_{eff}$ for all 625 pixels in the ROI, calculated from ODMR spectra using equation 1. While the distribution peaks close to zero at $B_{eff}$ = 0.5 Gauss without $Fe_2O_3$ particles, it significantly broadens and shifts to positive values with a peak at $B_{eff}$ = 1.0 Gauss. The shift shows the average effect of the magnetic field of the $Fe_2O_3$ particles on the NV ODMR spectra for the FNDs in the ROI. The spread is consistent with the fact that the local magnetic field within the ROI is expected to vary significantly.

Figure 3f) shows a map of $B_{eff}$ for the entire field of view shown in Figure 3a), where the regions covered by $Fe_2O_3$ particles are shown in white. The map shows the strongest local effective magnetic field in the areas surrounding the largest $Fe_2O_3$ particle aggregate, with a maximum $B_{eff}$ value of 3.5 Gauss, which decreases to values close to zero at 10 – 20 μm distance from the aggregate. These values align with previously reported measurements for similar particles,[34] given that $B_{eff}$ is a relative parameter.

While large $Fe_2O_3$ particles are expected to generate a constant local magnetic field, smaller isolated particles create magnetic noise. Figure 3c) shows NV $T_1$ decays with and without $Fe_2O_3$ particles deposited on the FND layer, acquired from the same ROI as the ODMR spectra. In the absence of $Fe_2O_3$ particles, the FNDs show a typical $T_1$ decay time of 600 μs, which reduces to 80 μs in the presence of magnetic particles as determined via a single exponential fit to the decay. Figure 3e) shows a histogram of the $T_1$ decay times for all pixels in the ROI. The peak of the distribution decreases by approximately an order of magnitude from 600 μs (no $Fe_2O_3$) to ~60 μs (with $Fe_2O_3$ particles). Similar to the results for $B_{eff}$, the $T_1$ values of FNDs in the ROI that are further spatially separated from the $Fe_2O_3$ particles only decrease slightly.

Figure 3g) shows a map of the $T_1$ decay time for the entire FOV. The map shows the most pronounced decrease in $T_1$ times occurring in the vicinity of smaller $Fe_2O_3$ particles scattered throughout the top half of the image. In the grey areas, we were unable to determine decay times based on exponential fits, likely due to strong local magnetic noise reducing the decay time below our measurement resolution of 10 μs. The significant reduction in $T_1$ decay times observed across the FOV is caused by the presence of magnetic noise generated by the fluctuating magnetic moments of $Fe_2O_3$ nanoparticles. Many regions near the large $Fe_2O_3$ particle aggregate in the bottom left quadrant of the image, on the other hand, show only a small reduction in NV $T_1$ decay times, in agreement with the static effective magnetic fields measured via ODMR. The existence of both static and fluctuating magnetic fields is expected for this sample as the size of the iron oxide particles determines whether they exhibit superparamagnetic behavior instead of weak ferromagnetism. Similar results have been shown in prior NV sensing work[34]. The results presented in Figure 3 show that FND layers self-assembled on quartz and glass substrates can be used to image static magnetic fields and magnetic noise on the micrometer scale.

The benefits of these monolayer substrates are two-fold; the first benefit is that a single layer of FNDs ensures that the interaction between the target of interest (here: the $Fe_2O_3$ particles) and NVs is maximized, as there is no background contribution to the signal from particles that are further spatially separated in the z-direction. Secondly, the substrates can be used for imaging and sensing applications as a cost-efficient and scalable – both in number and in size – alternative to single-crystal diamond, which is currently typically used for such applications. The FND layers reported here make mass-produced and disposable quantum sensors based on nanodiamonds feasible.

In summary, we have presented a simple and scalable method for creating FND layers on positively charged surfaces and demonstrated that these layers are useful for magnetic imaging using NV PL. We have investigated the effect of FND concentration in suspension, substrate immersion time, and pH and ionic strength of the solvent water on the FND density on the substrate surface. We find that an FND concentration of 1 mg mL$^{-1}$ and an immersion time of 1000 s (~16 min) without the addition of HCl optimizes the density of single FNDs on the substrate surface while avoiding significant aggregation. Longer immersion times will increase the total FND density on the surface (single FNDs and aggregates), but likely at the expense of more aggregates. By depositing magnetic $Fe_2O_3$ particles on the FND layers on a quartz substrate, we demonstrated that quantum diamond microscopy can be used to image effective magnetic fields and magnetic noise on the micrometer scale, based on NV ODMR magnetometry and NV $T_1$ spin relaxometry, respectively. We showed that static magnetic fields can be distinguished from magnetic noise. Our results pave the way for the development of functional FND layers and coatings for on-demand NV-based quantum sensing and imaging on a wide range of substrates.

**Acknowledgements**
PR acknowledges support through an Australian Research Council (ARC) DECRA Fellowship (grant no. DE200100279), ARC Discovery Projects (DP220102518, DP250100125), and an RMIT University Vice-Chancellor's Senior Research Fellowship. J-PT acknowledges support through an ARC Future Fellowship (FT200100073). This work was performed in part at the RMIT Micro Nano Research Facility in the Victorian node of the Australian National Fabrication Facility (ANFF) and the RMIT Microscopy and Microanalysis Facility (RMMF).

**Methods**
*Materials*. Nanodiamonds made from high-pressure high-temperature diamonds were purchased from Nabond, China (120 nm nominal particle size), irradiated with 2 MeV electrons to a fluence of $1 \times 10^{18}$ e$^-$ cm$^{-2}$, annealed in an argon atmosphere at 900°C for 2 h, oxidized in air for 2.5h, and dispersed in DI water at 10 mg mL$^{-1}$ using bath sonication. Larger All FND suspensions investigated for FND self-assembly were made from this stock suspension. HCl, NaCl and polyallylamine hydrochloride (PAH) (all from Merck) were used as received. P-type Si wafers (Merck) and quartz substrates (Lanno Quartz, China, 0.5 mm thickness) were cut into $5 \times 5$ mm substrates for self-assembly experiments.

*FND self-assembly*. The substrates were cleaned using sonication in acetone, ethanol and DI water, followed by UV ozone cleaning (10 min). The substrates were then vertically immersed in an aqueous PAH solution (10 mg mL$^{-1}$, 5 min), rinsed with DI water, and dried under nitrogen flow. The substrates were vertically immersed in different FND suspensions for different times as discussed in the main text. After immersion in FND suspensions, the substrate is rinsed with DI water and dried on a hotplate at 100 °C (10 min).

*AFM imaging*. AFM images were acquired using an Asylum MFP-3D Infinity Atomic Force Microscope (Oxford Instruments, UK) in tapping mode with a Tap150AI tip (BudgetSensors, Bulgaria). A typical $50 \times 50$ μm image had $1024 \times 1024$ pixels with a scan rate of 0.25 Hz. See supporting information for full-sized images.

*PL imaging*. The image in Figure 1d) was acquired with a commercial wide field fluorescence microscope (IX83, Olympus, Japan), using 560 nm excitation, a 20× objective, a custom filter cube (570 nm dichroic and 600 nm long pass filter), and a Hamamatsu ORCA-Flash4.0 CMOS camera (Hamamatsu Photonics, Japan). Confocal PL images were acquired using a custom-built microscope. 532 nm laser (200 μW) excitation was focused onto the sample with a 100× air objective (0.9 NA), raster scanned across the sample piezo scanning stage (PInanoXYZ, PhysikInstrumente, Germany),

PL collected with the same objective, separated from the excitation signal with 532 nm dichroic and long pass filters, and detected with a avalanche photodiode (SPCM-AQRH-14, Excelitas Technologies, USA).

*Widefield ODMR and $T_1$ measurements* were carried out on a custom-built wide-field fluorescence microscope. Optical excitation was provided by a pulsed wave (PW) laser at 532 nm (Laser Quantum Opus 2 W, Lastek, Australia), using a widefield lens (125 nm) the beam was focused to the back aperture of the objective lens (S Plan Fluor ELWD 20×, NA = 0.45, Nikon, Japan). The laser intensity at the sample was less than 500 mW in all cases. Fluorescence from NV centers was separated from the excitation beam with a dichroic mirror (538 nm brightline dichroic beamsplitter, Semrock, United States), a 650 nm long-pass filter (ThorLabs, United States), and an 850 nm short-pass filter and collected with a CMOS camera (Zyla 5.5-W USB3, Oxford Instruments, UK). Microwave (MW) driving was provided by a signal generator (Synth NV pro, Windfreak Technologies, United States) gated using an IQ modulator (TRF37T05EVM, Texas Instruments, United States) and amplified with a (HPA-50W-63+, Mini-Circuits, United States). A pulse pattern generator (PulseBlasterESR-PRO 500 MHz, SpinCore, United States) was used to modulate the excitation laser and MW and to synchronize the image acquisition. MWs were delivered to the samples via a stripline waveguide in a printed circuit board (PCB). Samples were positioned on top of the exposed stripline.

Supporting Information for

# Self-assembled fluorescent nanodiamond layers for quantum imaging

Katherine Chea[1], Erin S. Grant[1], Kevin J. Rietwyk[1], Hiroshi Abe[2], Takeshi Ohshima[2,3], David A. Broadway[1], Jean-Philippe Tetienne[1], Gary Bryant[1], Philipp Reineck[1]

[1] *School of Science, RMIT University, Melbourne, VIC 3001, Australia*
[2] *National Institutes for Quantum and Radiological Science and Technology, Takasaki, Gunma, 370-1292, Japan*
[3] *Department of Materials Science, Tohoku University, Aoba, Sendai, Miyagi 980-8579, Japan*

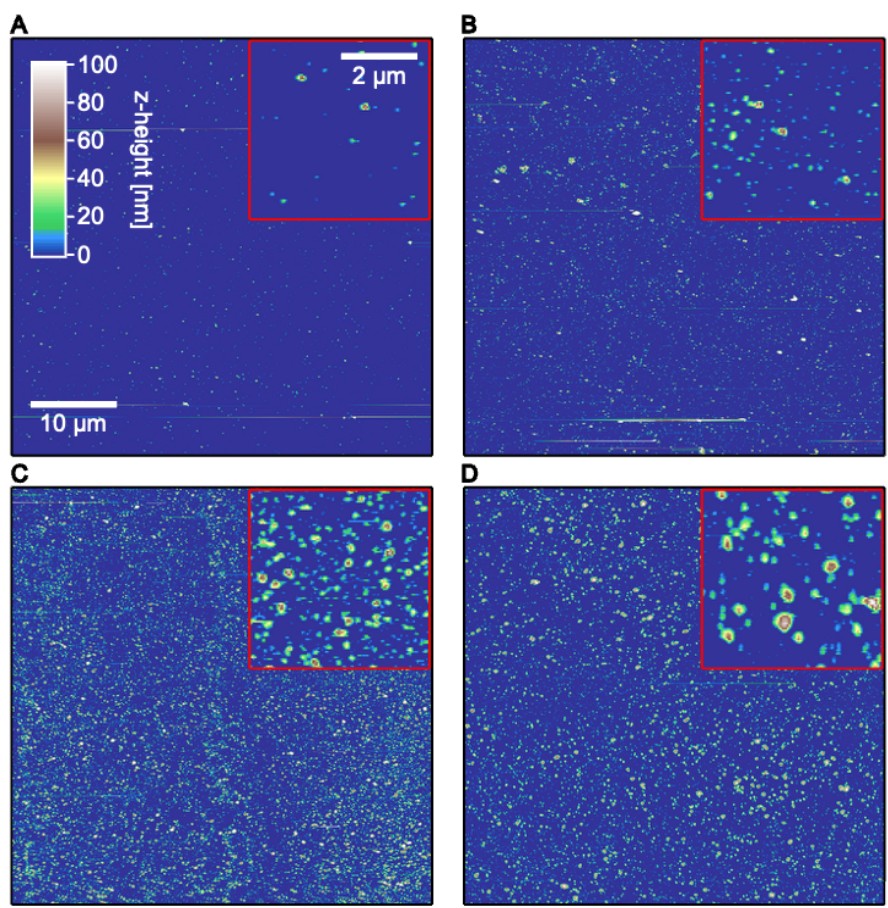

**Figure S1**. 50 µm × 50 µm images of FNDs on a Si wafer substrate. The FND layers were created using FND concentrations of 0.01 (a) 0.1 (b), 1, (c) and 10 mg mL$^{-1}$ in suspension for an immersion time of 100 s. The inset in the top right corner of each image shows a zoomed in region. The scale bars shown in panel a) apply to all images.

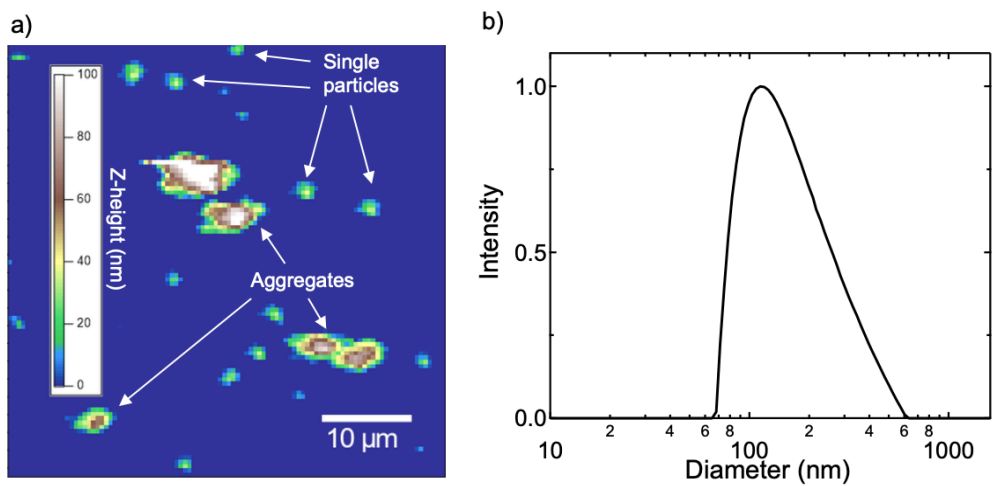

**Figure S2.** a) AFM image showing single FND particles and aggregates. By analyzing line profiles of the z-height of the features shown in this image, we identified a z-height of 55 nm as a threshold that separates single FNDs from larger aggregates in most instances. By using a z-height-based threshold, we also count multiple FNDs that are located in close vicinity to one another in the x-y plane, but are not stacked, as 'single particles'. b) Unweighted dynamic light scattering particle size distribution peaking at a diameter of $D = 120$ nm.

## Surface to surface distance calculation

We assumed a spherical disk particle shape with diameter $D=120$ nm and calculated the average surface-to surface distance using

$$d_{avg} = \sqrt{\frac{A_{tot}}{N_{FND}}} - D$$

Where $A_{tot}$ is the total surface area, $N_{FND}$ is the number of FNDs attached to this area.

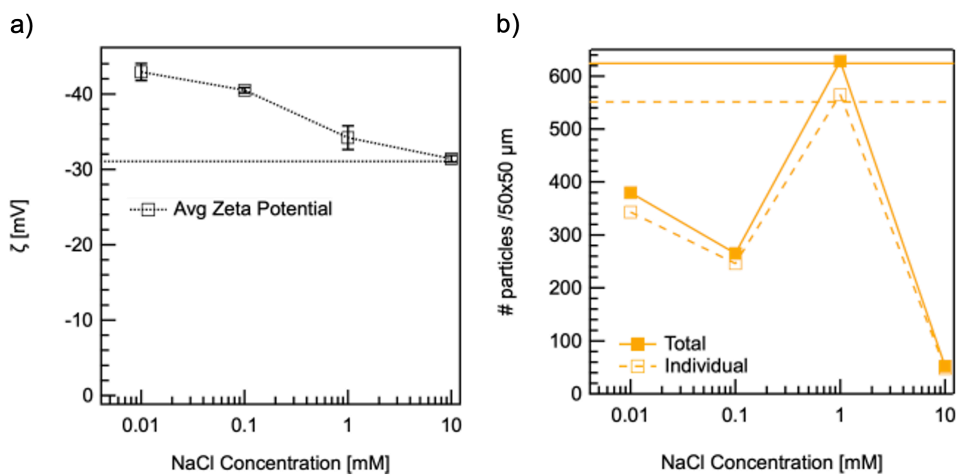

**Figure S3.** FND zeta potential (a) and number of FNDs on a 50 × 50 μm substrate area as a function of NaCl concentration in suspension. The dotted horizontal line in a) indicates the zeta potential without any added salt. The solid and dashed traces in b) show the total number of FNDs and the number of single FNDs on the substrate surface respectively.

**Table S1.** The measured pH of FND suspensions with varying HCl concentration compared to the calculated pH values.

| HCl conc [μM] | Calculated pH | Measured pH |
|---|---|---|
| 0 | 7 | 6.4 |
| 1 | 6 | 6 |
| 10 | 5 | 5.8 |
| 100 | 4 | 5 |
| 1000 | 3 | 3.8 |

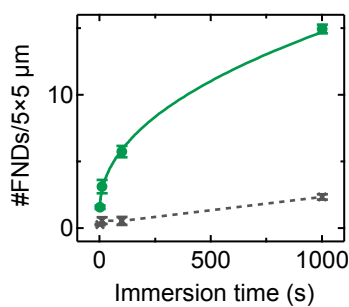

**Figure S4**. The number of FNDs on a 50 × 50 μm substrate area as a function of substrate immersion time with a linear x-axis scaling. The solid green and dashed grey traces show single FNDs and aggregates, respectively. The green trace is a square root immersion time fit to the experimental data (markers).

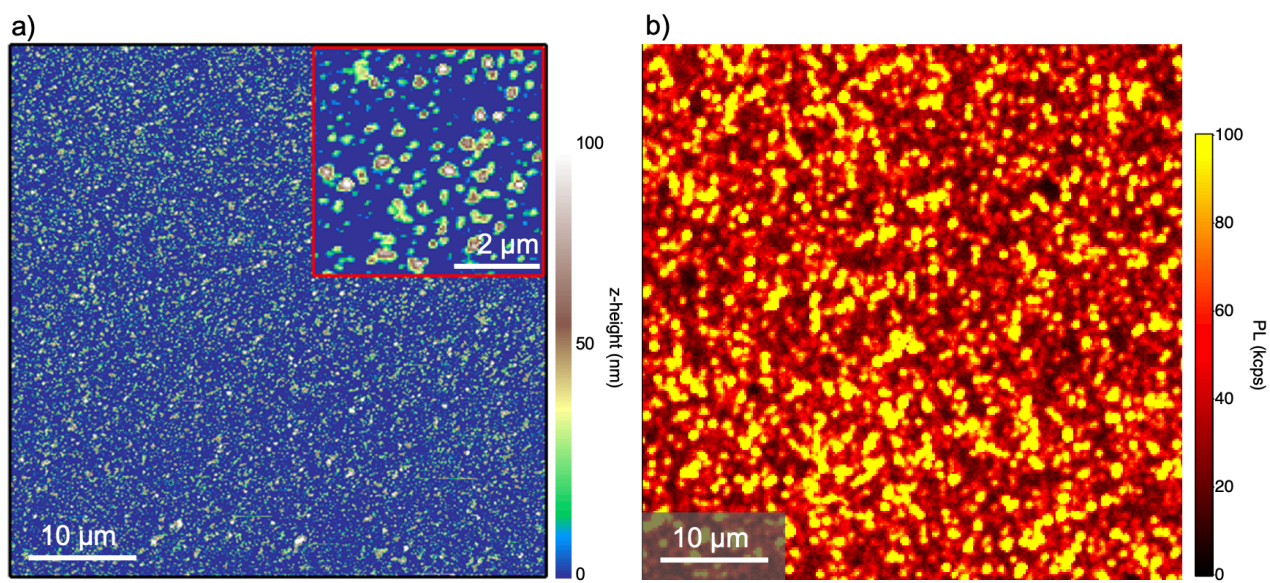

**Figure S5**. AFM (a) and PL (b) images of a high-density FND layer on a Si wafer. Both images are plotted at the same scale for direct comparison. Optimized self-assembly parameters of 1 mg mL$^{-1}$ FNDs and 1000 s substrate immersion time were used to fabricate the sample. Based on a z-height threshold of 10 nm, an FND surface coverage of 16.8 % was determined from the AFM images. Using a PL intensity threshold of 20 kcps, which is one more than an order of magnitude above the dark counts of the avalanche photodiode used for imaging, an NV signal surface coverage of 91.4% was determined. For an intensity threshold of 10 kcps, this coverage increases to 99.9%. The PL image in b) is plotted strongly saturated to visually demonstrate that FNDs are present in the entire field of view. The brightest FNDs show PL intensities above 1 Mcps.

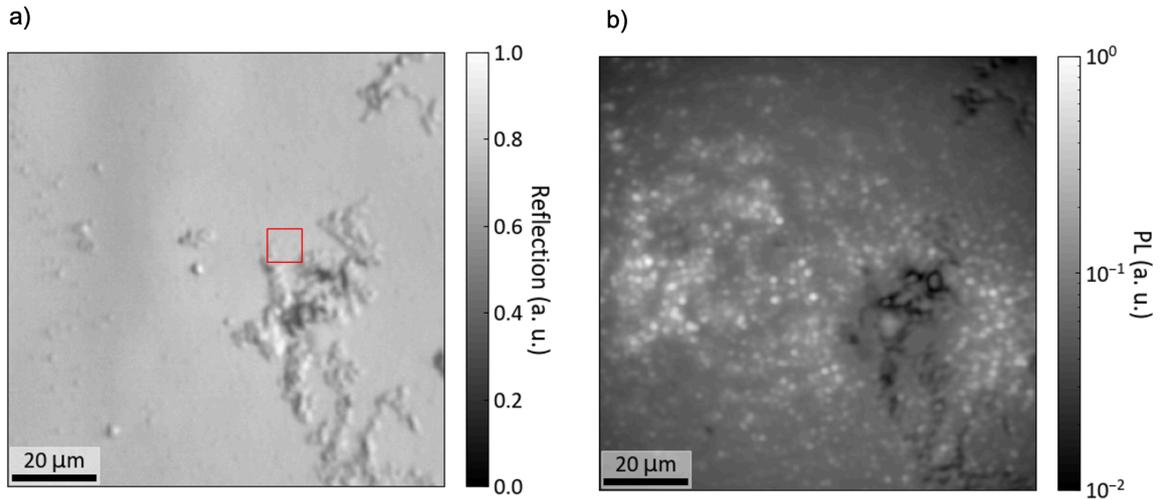

**Figure S6**. a) Bright field image of the Fe$_2$O$_3$ particles on the quartz substrate rotated 90° anticlockwise relative to the image in the main text Figure 3a. The red box indicates the region where the 'with Fe$_2$O$_3$' ODMR spectrum and T1 decay trace in the main text Figures 3b) and 3c), respectively, were acquired. b) Wide-field PL image of the same field of view as in a), showing FND PL across most of the field of view. The PL intensity is strongly reduced in the area covered by the Fe$_2$O$_3$ particles, which absorb and scatter both green excitation light and NV PL.

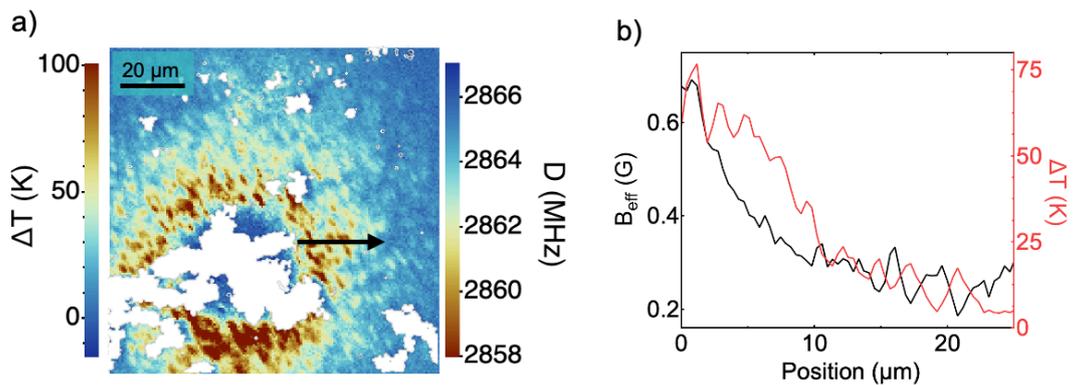

**Figure S7**. a) Map of the local change in temperature (left color scale), determined from the shift in the ODMR resonance frequency (right color scale). We used a frequency shift per Kelvin temperature difference of 74 kHz K$^{-1}$.[1]
b) Effective local magnetic field B$_{eff}$ and change in temperature ΔT as a function of distance from the large aggregate, as indicated by the black arrow in panel a). B$_{eff}$ decreases much faster than the temperature and the signals are not correlated, demonstrating that the magnetic and temperature signals can be separated.

**References**

[1]. V.M. Acosta, E. Bauch, M.P. Ledbetter, A. Waxman, L.-S. Bouchard, D. Budker, *Phys.Rev.Lett*. **2010**,104,070801.